# A Review of Software Quality Models for the Evaluation of Software Products

José P. Miguel[1] , David Mauricio[2] and Glen Rodríguez[3]

[1]Department of Exact Sciences, Faculty of Sciences, Universidad Peruana Cayetano Heredia, Lima, Peru
jose.miguel@upch.pe

[2]Faculty of System Engineering and Computing, National University of San Marcos, Lima, Peru
dms_research@yahoo.com

[3]Faculty of Industrial and System Engineering,  National University of Engineering, Lima, Peru
glen.rodriguez@gmail.com

*Abstract*

*Actually, software products are increasing in a fast way and are used in almost all activities of human life. Consequently measuring and evaluating the quality of a software product has become a critical task for many companies. Several models have been proposed to help diverse types of users with quality issues. The development of techniques for building software has influenced the creation of models to assess the quality. Since 2000 the construction of software started to depend on generated or manufactured components and gave rise to new challenges for assessing quality. These components introduce new concepts such as configurability, reusability, availability, better quality and lower cost. Consequently the models are classified in basic models which were developed until 2000, and those based on components called tailored quality models. The purpose of this article is to describe the main models with their strengths and point out some deficiencies. In this work, we conclude that in the present age, aspects of communications play an important factor in the quality of software.*

## 1. Introduction

Research on software quality is as old as software construction and the concern for quality products arises with the design of error-free programs as well as efficiency when used. Research to improve the quality of software is generated due to users demand for software products with increasing quality. Actually, this is considered an engineering discipline [1].

According to the IEEE Standard Glossary of Software Engineering Terminology [2,3,28], the quality of software products is defined as 1) the degree to which a system, component or process meets specified requirements and 2) the degree to which a system, component or process meets the needs or expectations of a user.

An acceptable definition for a software product, given by Xu [4], was "a packaged software component configuration or a software-based service that may have auxiliary components and which is released and exchanged in a specific market". Here packaged components refer to all kinds of programs. The software product takes different forms [4]: small, COTS (Commercial Off-The-Shelf Components), packed software, large commercial software, open source software and services

In this paper we focus on the quality of the software product, that is, in the final product rather than on the processes that lead to its construction, even though they are closely related.

The use of models is an acceptable means to support quality management software products. According to ISO/IEC IS 9126-1  [5] a quality model is "the set of characteristics, and the relationships between them that provides the basis for specifying quality requirements and evaluation". The models to evaluate the quality of software have been constructed defining the fundamental factors (also called characteristics), and within each of them the sub factors (or sub characteristics). Metrics are assigned to each sub factor for the real evaluation.

Figure 1 updates the work of Thapar [6] and shows the evolution of quality models from the Mc Call first model in 1977 until 2013. This evolution has categorized the models in: the Basic Models (1977 - 2001) whose objective is the total and comprehensive product evaluation [6] and the Tailored Quality Models (from 2001 onwards) oriented to evaluations of components. In this work models oriented to evaluation of Free Software are also considered because of their actual importance.

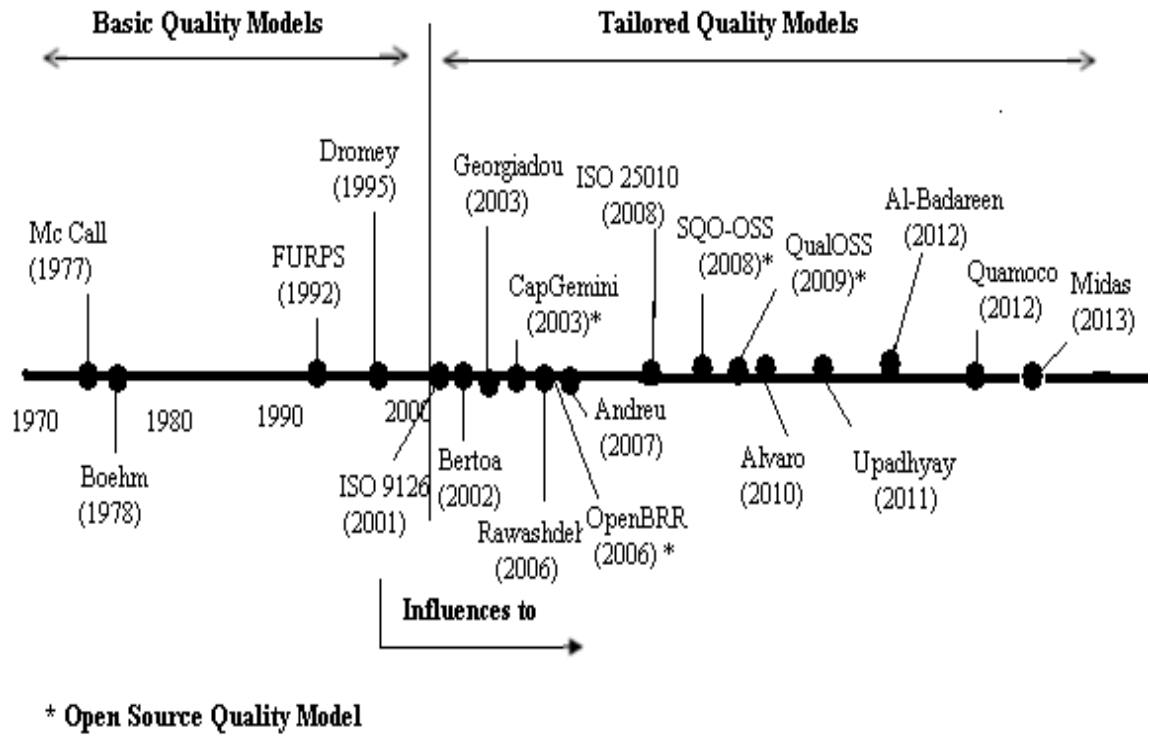

Figure 1 Quality Models

The Basic Models are hierarchical in structure; they can be adjusted to any type of software product and are oriented to the evaluation and improvement. The six most important are: Mc Call et al in 1977 [7], Boehm et al in 1978 [8], FURPS Model in1992 [9], Dromey model in 1995 [10], ISO 9126-1 model in 2001 [5] and its standards for both external metrics: ISO / IEC 9126-2 in 2003 [11], internal metrics: ISO / IEC 9126-3 in 2003 [12] and quality in use: ISO / IEC 9126-4 in 2004 [13]. The ISO -9126 model received inputs from previous models and sets standards for assessing the quality of software. In 2007 an updated was established as the ISO 25010 model: ISO / IEC CD 25010 [14]. The ISO 25010 actually is known as SQuaRE (Software engineering- Software product Quality Requirements and Evaluation).

Tailored Quality Models began to appear the year 2001 with Bertoa model [15], followed by Georgiadou Model in 2003 [16], Alvaro Model in 2005 [17], Rawashdesh Model [18]. The main characteristic is that they are specific to a particular domain of application and the importance of features may be variable in relation to a general model. These models arise from the need of organizations and the software industry for specific quality models capable of doing specialized evaluation on individual components. They are built from the Basic Models, especially the ISO 9126, with the adding or modification of sub factors and the goal to meet needs of specific domains or specialized applications. In recent years the software construction has focused on the reuse and development of Component-Based Software (CBSD). As a consequence the success of a product strongly depends on the quality of the components.

Other authors classify the models according to user's characteristics. For example Klas [19] distinguishes three categories of models that correspond to: 1) the level of general public use or specific domain, 2) organizational level that focus on satisfying the interests of a specific organization, and 3) the level project that applies to a specific project to ensure quality.

Due to the importance of COTS components Ayala [20] establishes a process to select software components. It was based on observations and interviews with developers of COTS-based components. The study concludes with varying results. One of the findings was discovering the use of informal procedures to find, evaluate and choose components, and hence there exists the need for methods to do components selection and support tools to help in the evaluation.

Some companies have also developed their own quality models, like the FURPS model [9] already mentioned and set by Hewlett Packard. A recent work by Samarthyam is the MIDAS model (Method for Intensive Design assessments) [21] established by the company Siemens that is used for the design of software products in the industry, energy, Health and Infrastructure. A description of some particular models used in businesses may be found in Pensionwar [22] and quality modelling for software product lines in Trendowicz [23].

We notice that many efforts have been done for the development of software product quality models. Furthermore several authors have done reviews of the literature on quality models and they included some benchmarking. Among these works we can mention: Al-Badareen in 2011 [24], Dubey in 2012 [25], Al-Qutaish in 2010 [26], Ghayathri in 2013 [27] and Samadhiya in 2013 [28]. All these works refer to the Basic Quality Models. In this work we review the literature of software product quality models including the Basic Models and the Tailored Models and based on the ISO 25010 model we perform a comparative evaluation. Finally and because of the increasing importance we include a review of product-oriented models for Open/Free Software.

This paper is organized as follows: section 2 describes the methodology used and a common terminology, shown in Table 1 is established, section 3 describes the Basic Quality Models, Section 4 describes some Tailored Quality Models according to their relevance, section 5 considers the Free Software oriented models, in Section 6 we make a comparative assessment of the models and in Section 7 some conclusions are established.

## 2. Methodology

### 2.1 Search strategies

Quality models have been found using the search engine Google Scholar, databases Science Direct, Ebsco, Trove (repository of information of the National Library of Australia) and NDTLD (Networked Digital Library of Theses and dissertations).

The main keywords used were "quality of software", "models for quality of software", "Evaluation of the quality of software", "metrics for evaluation of software", "general quality software product models" , "models for COTS components", "Models for free/open source quality", "Tailored quality models". The articles were classified according to the division established: Basic Quality, Tailored Models and Open Source Models.

### 2.2 Inclusion and exclusion criteria

The articles were classified according to their relevance preferring those describing models. In the state of the art articles we found several synonymous terms. Table 1 was constructed, using the literature review, to clarify the terminology and concepts related to quality. Regarding the exclusion criteria, the articles oriented to the evaluation of the software building process were set aside, since the purpose of the article is aimed at quality aspects of finished software products. The terminology mainly uses the international standards stated by the American Society for Quality [29] and in the ISO [5,11,12,13,14].

Table 1 Terminology used.

| Terminology | Synonyms | Definition | Reference |
|---|---|---|---|
| Acceptance | | Is all about the way the product is received in the user community, as this is largely indicative of the product's ability to grow and become a prominent product | (Duijnhouwer 2003) |
| Accountability | | The degree to which the actions of an entity can be traced uniquely to the entity. | (ISO/ IEC CD 25010 2008) |
| Accuracy | | The degree to which the software product provides the right or specified results with the needed degree of precision | (ISO/ IEC CD 25010 2008) |
| Adaptability | Versatility | The degree to which the software product can be adapted for different specified environments without applying actions or means other than those provided for this purpose for the software considered. | (ISO/ IEC CD 25010 2008) |
| Affordability | | How affordable is the component? | (Alvaro 2005) |
| Analyzability | | The degree to which the software product can be diagnosed for deficiencies or causes of failures in the software, or for the parts to be modified to be identified. | (ISO/ IEC CD 25010 2008) |
| Appropriateness | | The degree to which the software product provides an appropriate set of functions for specified tasks and user objectives. | (ISO/ IEC CD 25010 2008) |
| Appropriateness recognisability | Understandability | The degree to which the software product enables users to recognize whether the software is appropriate for their needs | (ISO/IEC 9126-1 2001), (ISO/ IEC CD 25010 2008) |
| Attractiveness | | The degree to which the software product is attractive to the user.. | (ISO/ IEC CD 25010 2008) |
| Authenticity | | The degree to which the identity of a subject or resource can be proved to be the one claimed | (ISO/ IEC CD 25010 2008) |
| Availability | | The degree to which a software component is operational and available when required for use. | (Dromey 1995) (ISO/ IEC CD 25010 2008) |
| Changeability | Changeable | The degree to which the software product enables a specified modification to be implemented. The ease with which a software product can be modified | (ISO/ IEC CD 25010 2008) |
| Co-existence | | The degree to which the software product can co-exist with other independent software in a common environment sharing common resources without any detrimental impacts | (ISO/ IEC CD 25010 2008) |
| Compatibility | | The ability of two or more software components to exchange information and/or to perform their required functions while sharing the same hardware or software. | (ISO/ IEC CD 25010 2008) |
| Confidentialit | | The degree to which the software product provides | (ISO/ IEC |
| y | | protection from unauthorized disclosure of data or information, whether accidental or deliberate. | CD 25010 2008) |
| Configurability | | The ability of the component to configurable. | (Alvaro 2005) |
| Compliance | Conformance | The degree to which the software product adheres to standards, conventions, style guides or regulations relating to a main factor. | (ISO/ IEC CD 25010 2008) |
| Correctness | | The ease with which minor defects can be corrected between major releases while the application or component is in use by its users | (Dromey 1995) |
| Ease of use | Usability, operability | The degree to which the software product makes it easy for users to operate and control it. | (ISO/IEC 9126-1 2001), (ISO/ IEC CD 25010 2008) |
| Efficiency | Performance Efficiency | The degree to which the software product provides appropriate performance, relative to the amount of resources used, under stated conditions | (IEEE 1993), (ISO/IEC 9126-1 2001), (ISO/ IEC CD 25010 2008) |
| Fault Tolerance | | The degree to which the software product can maintain a specified level of performance in cases of software faults or of infringement of its specified interface. | (ISO/IEC 9126-1 2001), (ISO/ IEC CD 25010 2008) |
| Flexibility | | Code possesses the characteristic modifiability to the extent that it facilitates the incorporation of changes, once the nature of the desired change has been determined. | (Ghayathri 2013) |
| Functionality | Functional suitability | The degree to which the software product provides functions that meet stated and implied needs when the software is used under specified conditions | (ISO/IEC 9126-1 2001), (ISO/ IEC CD 25010 2008) ASQ |
| Helpfulness | | The degree to which the software product provides help when users need assistance. | (ISO/ IEC CD 25010 2008) |
| Installability | | The degree to which the software product can be successfully installed and uninstalled in a specified environment. | (ISO/ IEC CD 25010 2008) |
| Integrity | | The degree to which the accuracy and completeness of assets are safeguarded. | (ISO/ IEC CD 25010 2008) |
| Interoperability | Compatibility | Attributes of software that bear on its ability to interact with specified systems. | (ISO/IEC 9126-1 2001), ASQ |
| Learnability | Easy to learn | The degree to which the software product enables users to learn its application. | (ISO/ IEC CD 25010 2008) |
| Maintainability | | The degree to which the software product can be modified. Modifications may include corrections, improvements or adaptation of the software to changes in environment, and in requirements and functional specifications | (ISO/ IEC CD 25010 2008) , (ISO/IEC 9126-1 |
| | | | 2001) |
| Modifiability | | Corrections, improvements or adaptations of the software to changes in environment and in requirements and functional specifications. | (IEEE 1998) ASQ |
| Modification Stability | | The degree to which the software product can avoid unexpected effects from modifications of the software | (ISO/ IEC CD 25010 2008) |
| Modularity | | The degree to which a system or computer program is composed of discrete components such that a change to one component has minimal impact on other components. | (ISO/ IEC CD 25010 2008) |
| Non-repudiation | | The degree to which actions or events can be proven to have taken place, so that the events or actions cannot be repudiated later. | (ISO/ IEC CD 25010 2008) |
| Performance efficiency | Performance | The degree to which the software product provides appropriate performance, relative to the amount of resources used, under stated conditions. | (ISO/ IEC CD 25010 2008) |
| Recoverability | Recovery | The degree to which the software product can re-establish a specified level of performance and recover the data directly affected in the case of a failure | (ISO/ IEC CD 25010 2008) |
| Reliability | | The degree to which the software product can maintain a specified level of performance when used under specified conditions. | (ISO/IEC 9126-1 2001), (ISO/ IEC CD 25010 2008) |
| Reusability | Adaptability | The degree to which an asset can be used in more than one software system, or in building other assets | (ISO/ IEC CD 25010 2008) |
| Replaceability | | The degree to which the software product can be used in place of another specified software product for the same purpose in the same environment. | (ISO/ IEC CD 25010 2008) |
| Resource utilization | . | The degree to which the software product uses appropriate amounts and types of resources when the software performs its function under stated conditions. | (ISO/ IEC CD 25010 2008) |
| Robustnesss | | The degree to which an executable work product continues to function properly under abnormal conditions or circumstances. | (Dromey 1995) (ISO/ IEC CD 25010 2008) |
| Scalability | | The ease with which an application or component can be modified to expand its existing capabilities. It includes the ability to accommodate major volumes of data. | (Dromey 1995) (Alvaro 2005) |
| Security | | The protection of system items from accidental or malicious access, use, modification, destruction, or disclosure | (ISO/ IEC CD 25010 2008) |
| Supportability | Support, adaptability | The ability to extend the program, adaptability and serviceability. The ease with which a system can be installed and the ease with which problems can be localized. | (Grady 1992). |
| Self-contained | | The function that the component performs must be fully performed within itself. | (Alvaro 2005) |
| Testability | | The degree to which the software product enables modified software to be validated | (ISO/ IEC CD 25010 2008) , |
| Technical | | The degree of operability of the software product | (ISO/ IEC |

| accessibility | | for users with specified disabilities. | (ISO/ IEC CD 25010 2008) |
|---|---|---|---|
| Time behaviour | | The degree to which the software product provides appropriate response and processing times and throughput rates when performing its function, under stated conditions. | (ISO/ IEC CD 25010 2008) |
| Transferability | Portability | The ease with which a system or component can be transferred from one environment to another (extend hardware or software environment). | (ISO/ IEC CD 25010 2008) ,(ISO/IEC 9126-1 2001) |

## 3. Basic quality models

According to their importance and following the timeline of figure 1, the main Basic models are described in this section. They are characterized because they make global assessments of a software product.

### 3.1 Mc Call Model

The Mc Call model established product quality through several features. These were grouped into three perspectives: Product Review (maintenance, flexibility, and testing), Product Operation (correct, reliable, efficient, integrity and usability) and Product Transition (portability, reusability and interoperability). Figure 2 shows the model.

The major contribution of the McCall method was to considerer relationships between quality characteristics and metrics. This model was used as base for the creation of others quality models [25].

The main drawback of the Call Mac model is the accuracy in the measurement of quality, as it is based on responses of Yes or No. Furthermore, the model does not consider the functionality so that the user's vision is diminished.

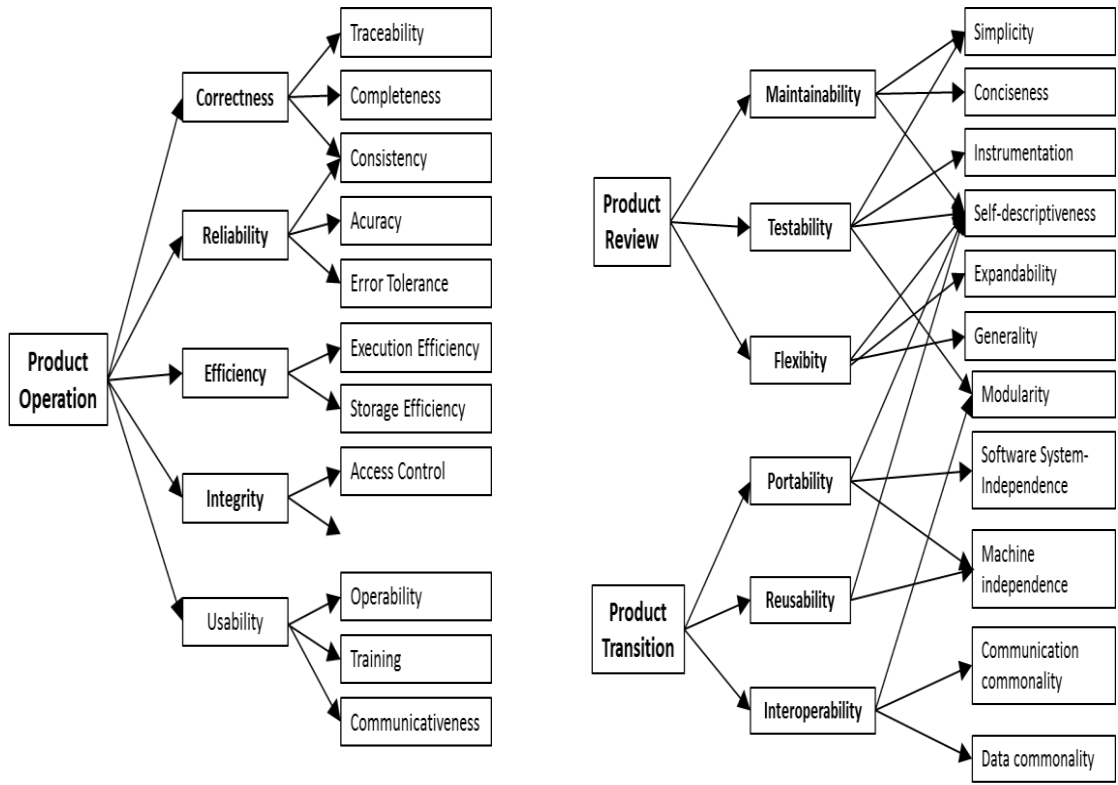

Figure 2 Mc Call Quality Model – 1977

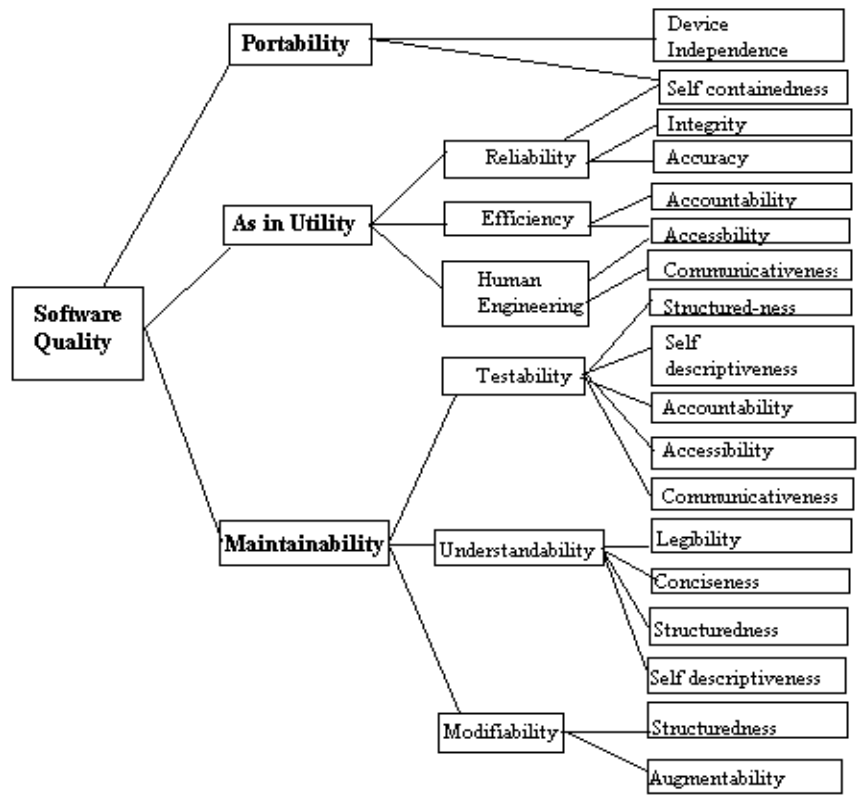

Figure 3 – Boehm Model -1978

### 3.2 Boehm Model

Boehm [8] establishes large-scale characteristics that constitute an improvement over the Mc Call model because adds factors at different levels. The high-level factors are: a) Utility indicating the easiness, reliability and efficiency of use of a software product;  b) maintainability that describe the facilities to modify,  the testability and the aspects of understanding;  c) portability in the sense of being able to continue being  used with a change of environment. Figure 3 [25] shows the model.

### 3.3 Dromey Model

The Dromey model [10] is based on the perspective of product quality. In this way the quality evaluation for each product is different and a more dynamic evaluation is established. The model states that for a good quality product, all the elements that constitute it, should be so. However, there is no discussion of how this can be done in practice, and this theoretical model is used to design others more specific models. Figure 4 shows the model.

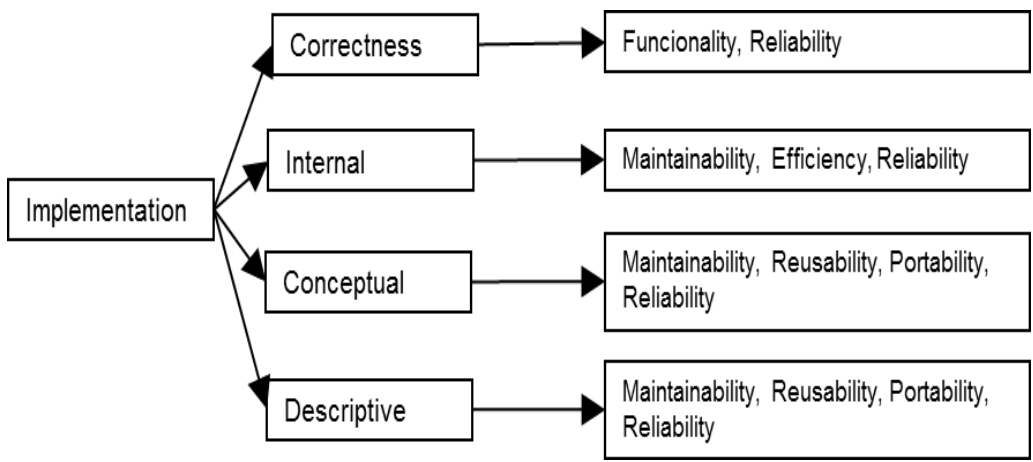

Figure 4 Dromey Model

### 3.4 FURPS Model

The model categorizes the characteristics as Functional Requirements (RF) and non-functional (NF). The RF are defined by the inputs and outputs expected or Functionality(F) while the NF are grouped as Usability (U), Reliability (R), Performance (P) and product support (S) [9]. Figure 5 shows these characteristics. Its main problem is that some main features, like portability, are not considered.

### 3.5 ISO 9126 Model

The ISO 9126 model [5] was based on the McCall and Boehm models. The model has two main parts consisting of: 1) the attributes of internal and external quality and 2) the quality in use attributes.

Internal quality attributes are referred to the system properties that can be evaluated without executing, while external refers to the system properties that can be assessed by observing during its execution. These properties are experienced by users when the system is in operation and also during maintenance.

The quality in use aspects are referred to the effectiveness of the product, productivity, security offered to the applications and satisfaction of users. Figure 6 [11,12] shows a view of the relationship between internal, external and quality in use attributes. Figure 7 and 8 illustrates the model [5,11,13].

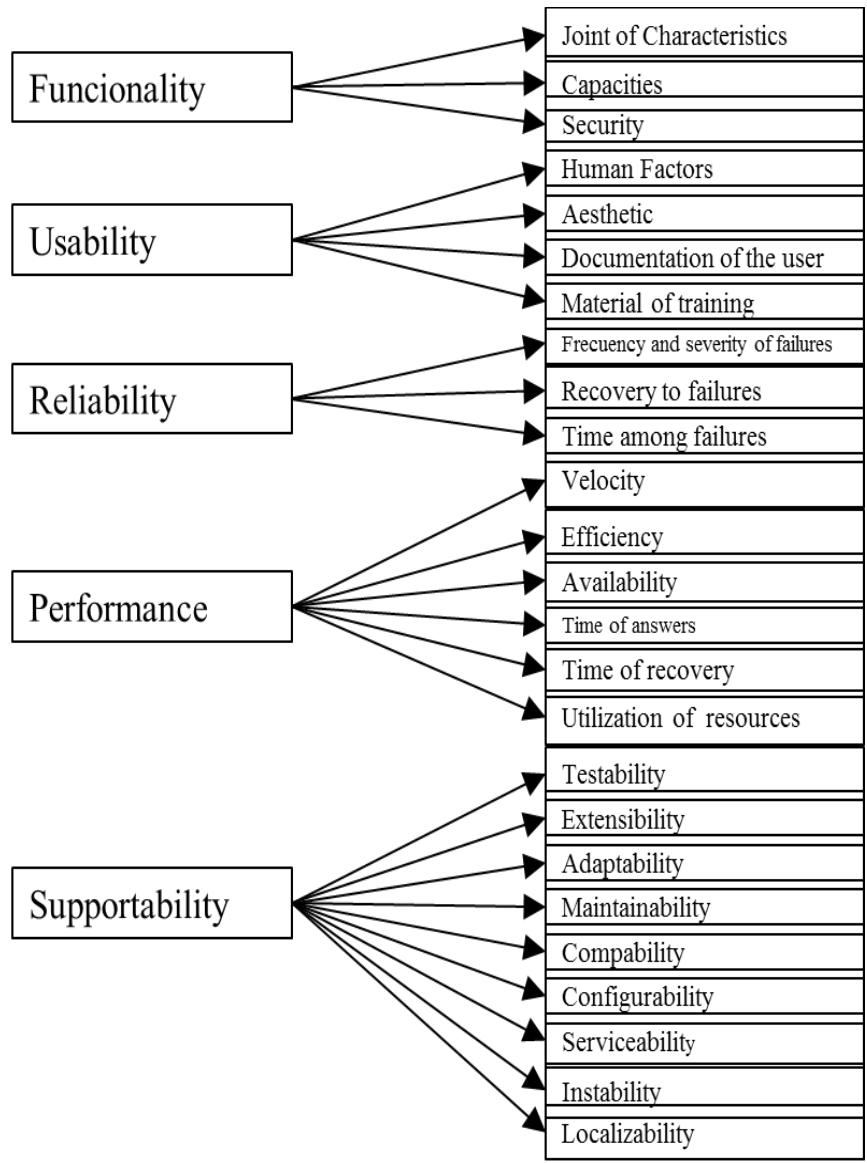

Figure 5 FURPS Model

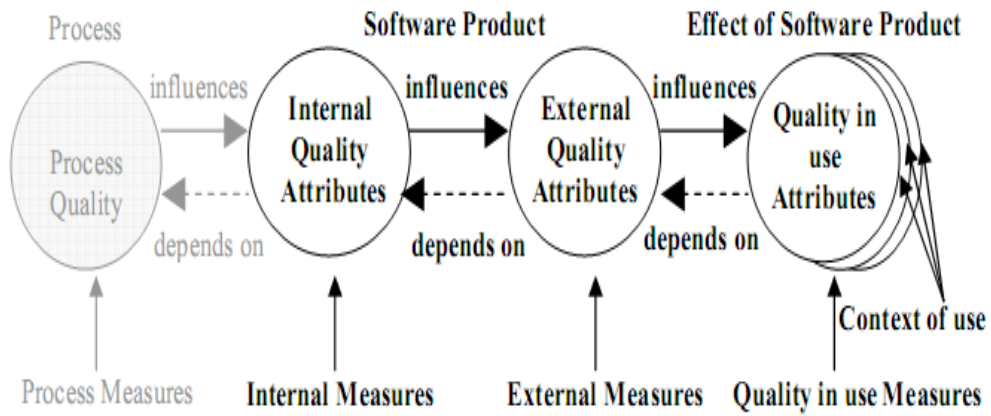

Fig. 6 Quality in the lifecycle ISO 9126

The ISO-9126 model has been used as the basis for Tailored Quality Models. One of its features was to standardize the terminology regarding quality of software.

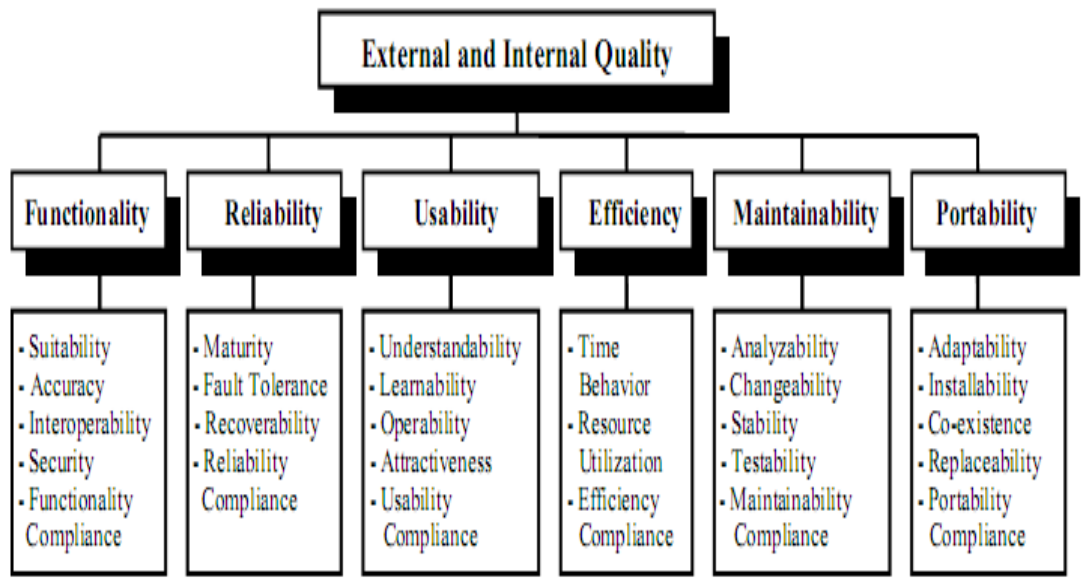

Figure 7. ISO 9126 Quality Model for external and internal quality

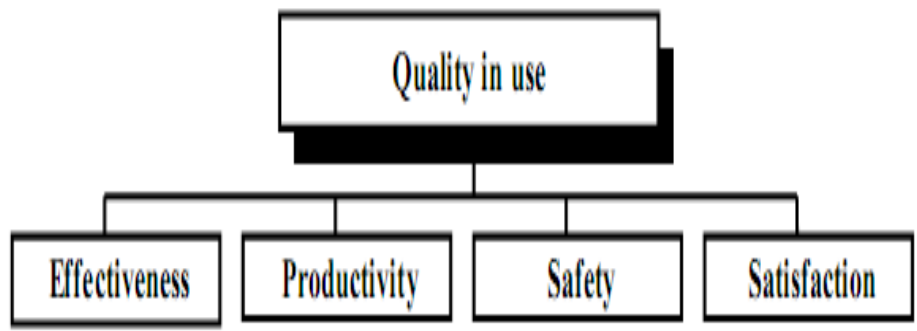

Figure 8. ISO 9126 Quality in use

### 3.6 ISO 25010 Model

This standard emerged in 2007 updating the ISO 9126 model. It is subdivided into 8 sub key features and characteristics. Constitute a set of standards based on ISO 9126 and one of its main objectives is to guide in the development of software products with the specification and evaluation of quality requirements. Figure 9 illustrates the model

This model considers as new characteristics the security and compatibility that groups some of the former characteristics of portability and those that were not logically part of the transfer from one environment to another. It uses the term transferability as an extension of portability.

As with the ISO / IEC 9126, this standard maintains the three different views in the study of the quality of a product, as they were illustrated in Figure 6 [14].

International Journal of Software Engineering & Applications (IJSEA), Vol.5, No.6, November 2014

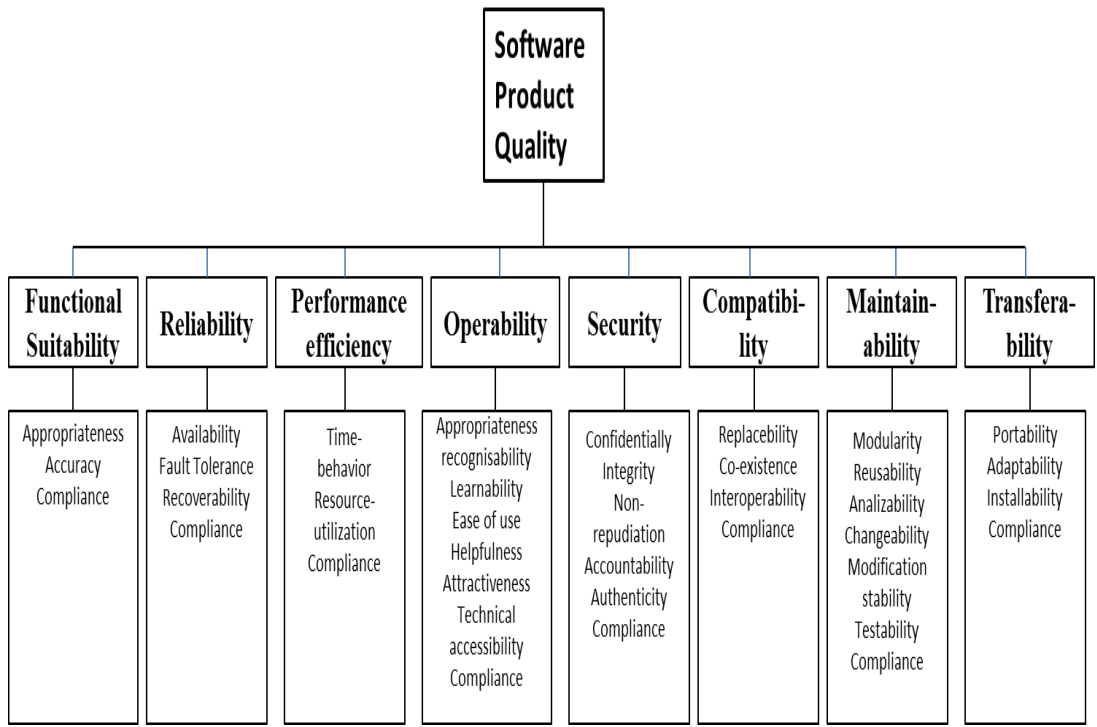

Figure 9 ISO 25010 Model (ISO/ IEC CD 25010 2007)

## 4. Tailored Quality Models

From 2001 the development of software was based on components (CBSD). The Non Basic models Software development concentrated on the use of Commercial Off-The-Shelf Components (COTS). Figure 10 illustrates the activities of the development of a product based on COTS available in the market

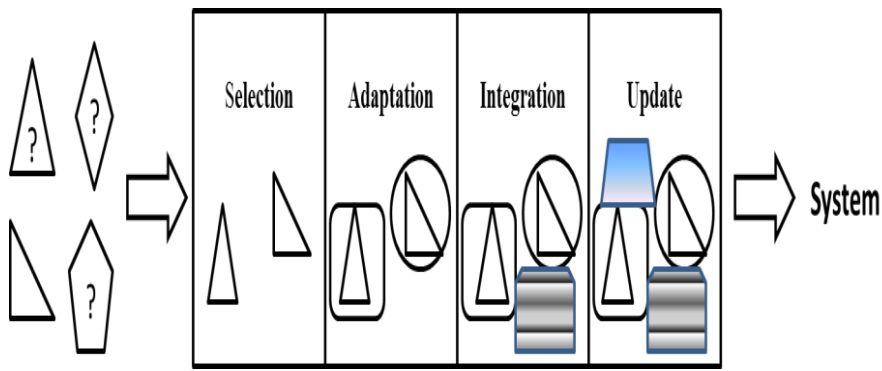

Figure 10 Activities for the construction of a System using components

**4.1 Bertoa Model**

The Quality Model Bertoa [15] is based on the ISO 9126 Model [5]. It defines a set of quality attributes for the effective evaluation of COTS. The COTS are used by software development companies to build more complex software. The model discriminates those features that make sense for individual components and is shown in figure 11.

International Journal of Software Engineering & Applications (IJSEA), Vol.5, No.6, November 2014

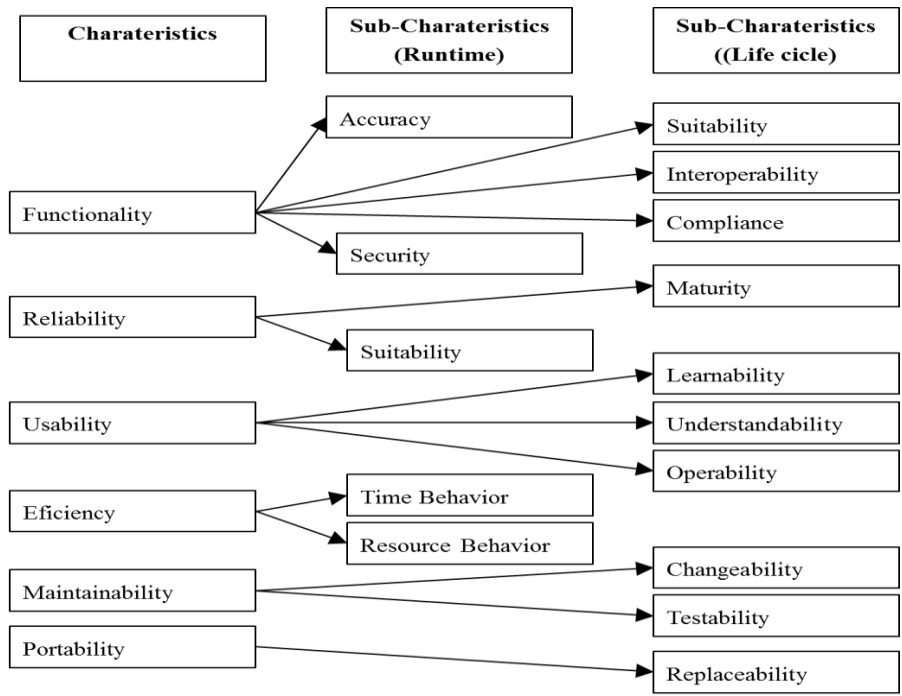

Figure 11  Bertoa Model

## 4.2 GEQUAMO

This model called GEQUAMO (Generic, Multilayered and Customizable Model), was created by E.Georgiadou [16] and consists of the gradual breakdown into sub layers of features and characteristics and is intended to encapsulate the various user requirements in a dynamic and flexible way. In this form the user (end user, developer, and manager) can build their own model reflecting the emphasis (weight) for each attribute and / or requirement. Figure 12 shows the decomposition of a CASE tool [16].

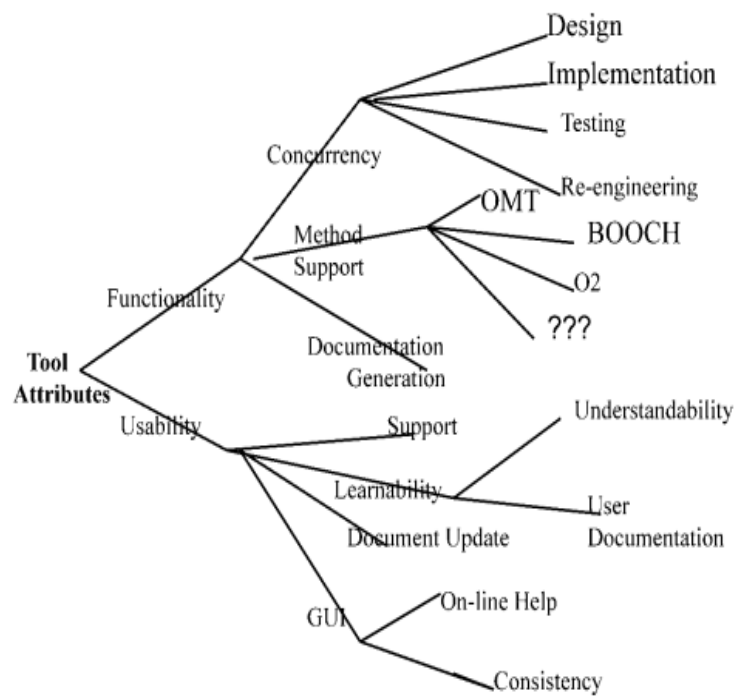

Figure 12 Layer of Characteristics applied to a tool CASE

### 4.3 Alvaro Model

Alvaro method considers a framework for the certification of software components) in order to establish the elements of quality components [17,30]. This framework considers four modules: 1) Model quality components for the purpose of determining the characteristics to be considered, 2) Framework for technical certification, which determines the techniques that will be used to evaluate the features provided by the model 3) the certification process that defines a set of techniques that evaluates and certifies the software components with the aim of establishing a well-defined component certification standard and 4) the frame containing the metric, which is responsible for defining a set of metrics evaluating the properties of the components in a controlled manner. In this article we refer to the quality components model. Figure 13 describes the model where the introduced sub-features are highlighted in bold.

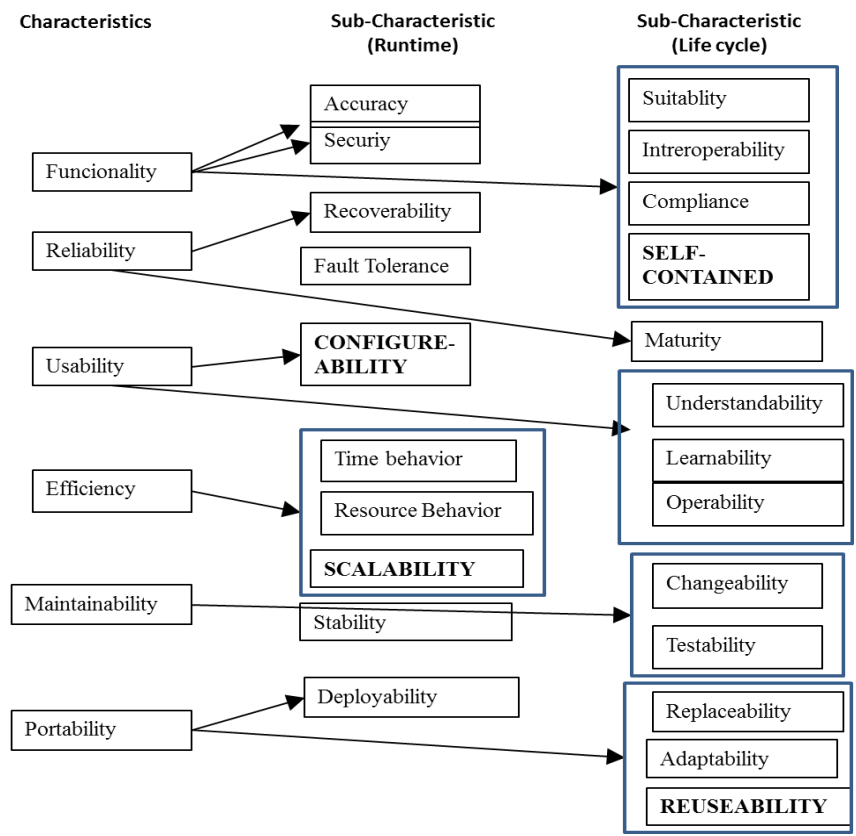

Figure 13.Alvaro Model

### 4.4 Rawashdeh Model

The Rawashdeh Model [18] has as main objective the needs of different types of users.

The model focuses on using components COTS and has been influenced by the ISO 9126 and Dromey models. The model sets out four steps to create a product quality model [18] that are:

- Identify a small group of high level quality attributes, then using a top down technique each attribute is decomposed into a set of subordinate attributes.
- Distinguish between internal and external metrics. Internal measure internal attributes such as specifications or source code, and external system behavior during testing operations and components.

- Identification of users for each quality attributes.
- Built the new model is with ideas of ISO 9126, and Dromey Model

Figure 14 shows the model.

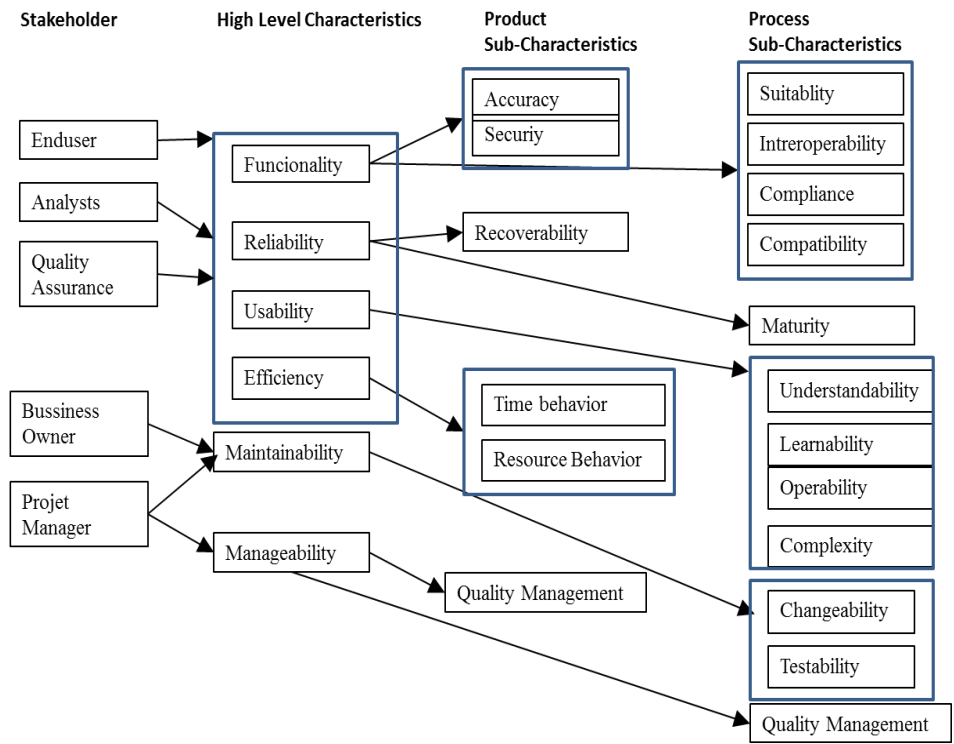

Figure 14– Rawashdeh Model

## 5. Open Source Models

Actually free Software products have much popularity for the diverse characteristics and freedoms they offer and because they are used in different contexts. Many of them are directed to perform the same or similar applications than traditional products. For example they can be Free Software Operating Systems (such as Linux, Solaris, FreeBSD), middleware technologies/databases (Apache Web Server, MySQL) and products for the end user (Mozilla Firefox, Open Office).

Models for assessing the quality of Free Software products adapt models like ISO-9126, adding some particular aspects of Free Software. It is noteworthy that although there is a distinction between models of first and second generation, an ideal model that captures all aspects of quality in a free software product has not been defined yet [31].

According to [32,33] these models started in 2003 and all of them emphasizes about the open source. In the next section we describe four models.

### 5.1 CapGemini Open Source Maturity Model

The model is based on the maturity of the product and is set according to maturity indicators. These indicators are grouped in product and application indicators [34]. For the final evaluation each of the sub indicators is given a value between 1 and 5 giving a total score. Figure 15 shows the model.

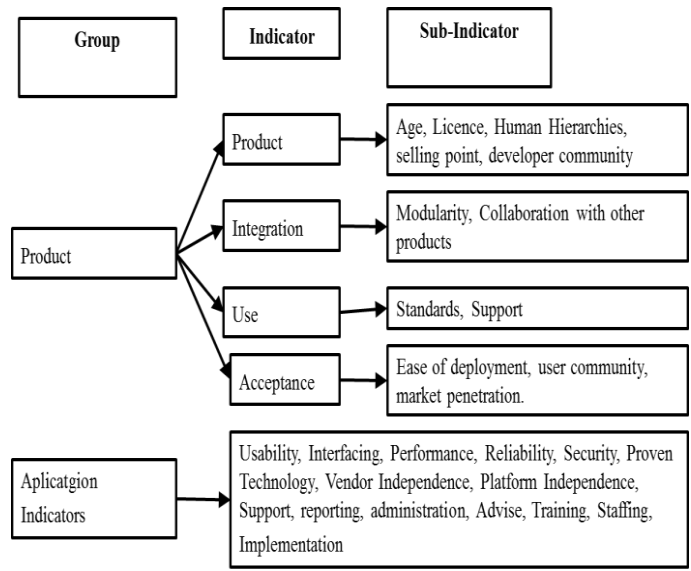

Figure 15  Cap Gemini Model for free/ open Software

## 5.2 OpenBRR Model

The model is called Business Readiness Rating framework and was influenced by the CapGemini and ISO 9126 Models. In this context identifies categories that are important for evaluating open software. The model has seven categories and thereby accelerates the evaluation process, ensuring better choices with a small set [32]. The seven categories can be refined for greater granularity and cover aspects that have not been considered at the highest level. The objective is to keep always in a very simple level [35].  Figure 16 shows the model.

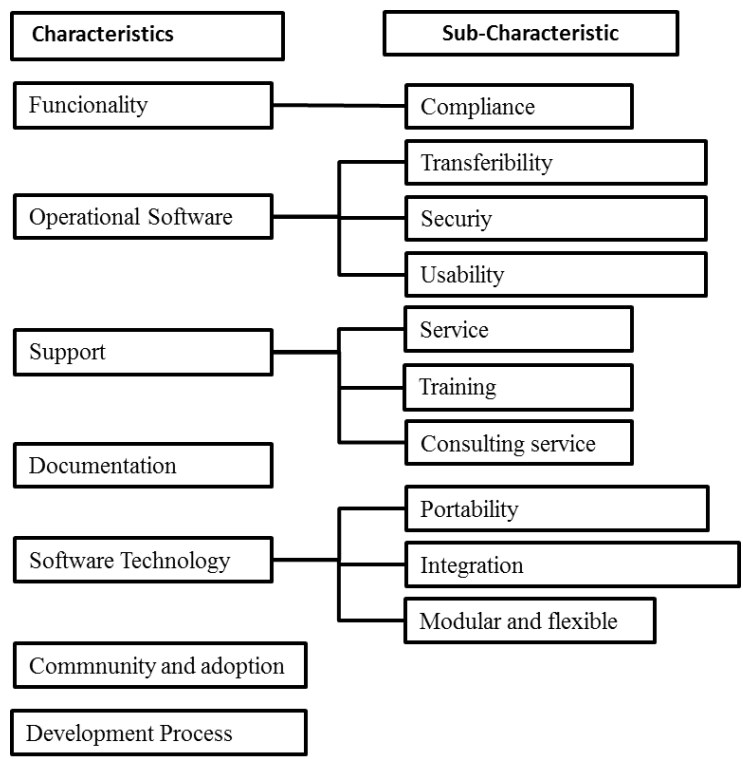

Figure 16  OpenBRR model

International Journal of Software Engineering & Applications (IJSEA), Vol.5, No.6, November 2014

**5.3 SQO-OSS Model.**

This is a hierarchical model that evaluates the source code and the community process allowing automatic calculation of metrics [32]. The model is show in figure 17 and according to [36], the model differs from others in the following aspects:

- Focus to the automation in contrast of other models that require heavy user interference.

- Is the core of a continuous quality monitoring system and allows automatic metrics collection.

- It does not evaluate functionality.

- It focuses in source code. Source code is the most important part of a software project.

- Considers only the community factors that can be automatically measured.

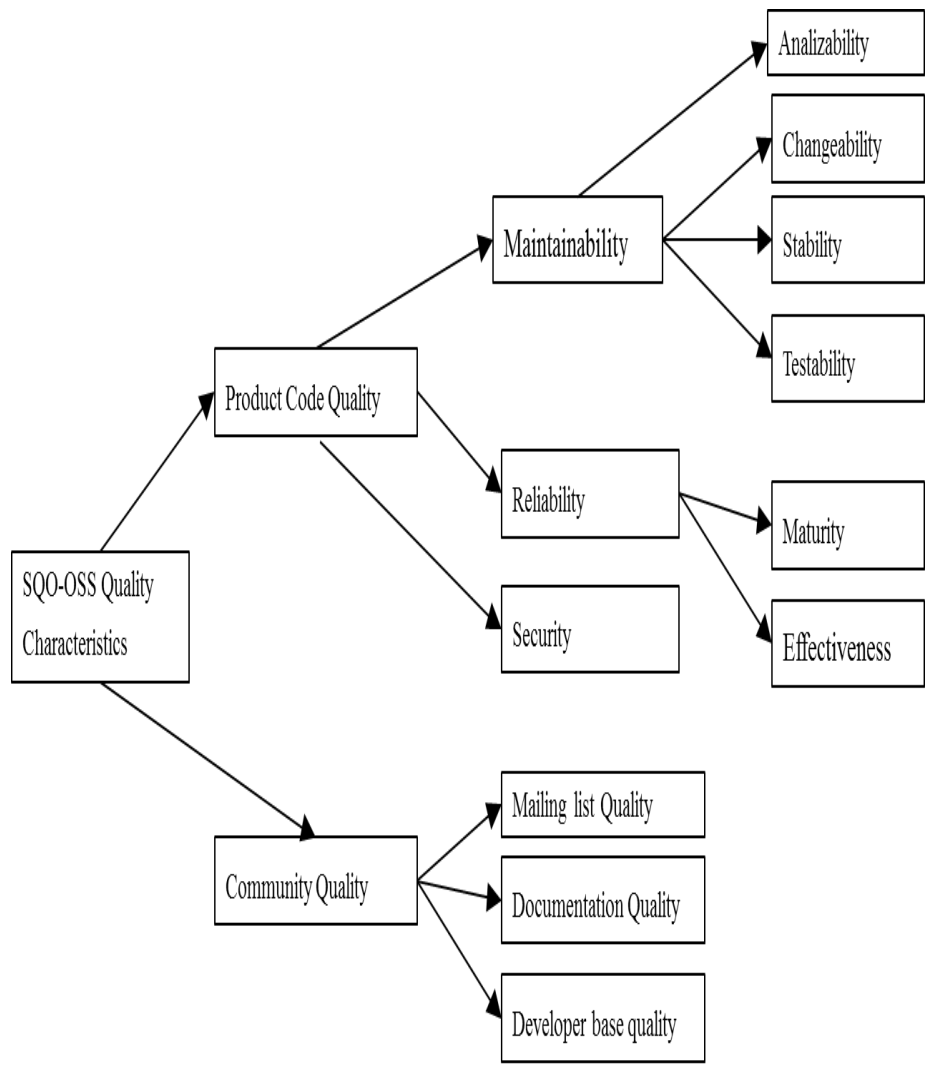

Figure 17 – SQO- OSS Model

**5.4 QualOSS Model**

It is a model that emphasizes three aspects: 1) Product characteristics, community characteristics and 3) Software process characteristics are equally important for the quality of a Free/ Open source product [33]. The model is shown in figure 18 [31].

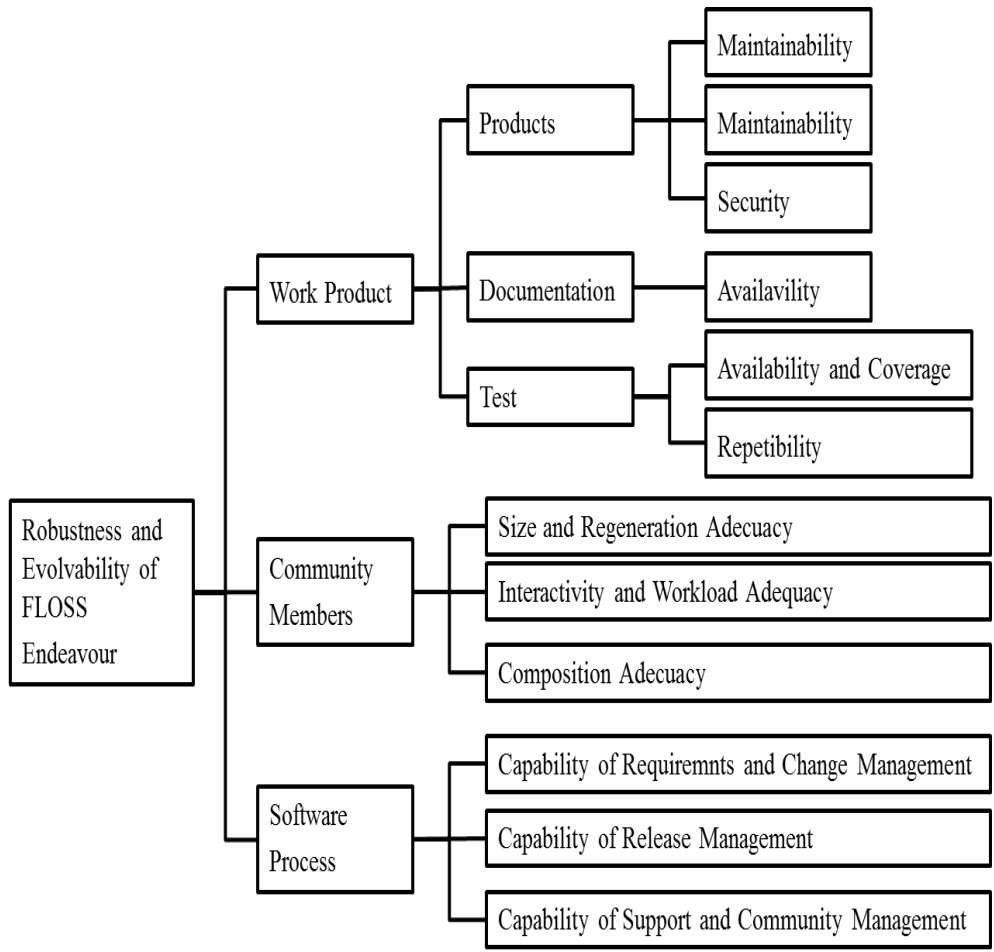

Figure 18 – QualOSS Model

The QualOSS model states that quality is highly depending on the context in which it is used an the purposes that a company or person pursues with it.

This model correspond to a second generation of Free/Open source models and where most of the assessment is highly automated.

## 6. Model Comparison

Al-Baradeen [24, 37], Al-Qutaish [25], Samarthyam [21] and Ghayathri [27] conducted comparative studies of Basics Quality Models, reaching different conclusions depending on the as they consider more important.

Table 2 shows a comparison of the basic models regarding the main characteristics according to Table 1. We include the ISO 25010 in this evaluation because it contains the last standardized terminology.

From table 2 we conclude that Model ISO 25010 is the most complete among the Basic Models, because it covers 26 of the 28 features. Flexibility is related to the manufacturing process [27] and is considered as an aspect of maintainability. Regarding Human Engineering this is a particular feature considered only in the Boehm model and has close relation with operability, but this last concept is wider.

From the table we conclude that reliability is a common feature to all models. The reason is the close relation with the opinion of users and the success of any product will depend on the fact of being used or not.

International Journal of Software Engineering & Applications (IJSEA), Vol.5, No.6, November 2014

Table 2 was constructed using the sub characteristics of the model. However and because these features are include in larger characteristic, it is possible that the presence of a feature implies that other has to be present. For example the transferability is related with some aspects of portability and adaptability.

Table 2 Comparison of Basic Models

| Characteristic | McCall | Boehm | FURPS | Dromey | ISO-9126 | ISO-25010 |
|---|---|---|---|---|---|---|
| Accuracy | | | | | X | X |
| Adaptability | | | X | | | X |
| Analyzability | | | | | X | X |
| Attractiveness | | | | | X | X |
| Changeability | | | | | X | X |
| Correctness | X | | | | | X |
| Efficiency | X | X | | X | X | X |
| Flexibility | X | | | | | |
| Functionality | | | X | X | X | X |
| Human Engineering | | X | | | | |
| Installability | | | | | X | X |
| Integrity | X | | | | | X |
| Interoperability | X | | | | | X |
| Maintainability | X | | | X | X | X |
| Maturity | | | | | X | X |
| Modifiability | | | | | | X |
| Operability | | | | | X | X |
| Performance | | | X | | X | X |
| Portability | X | X | | X | X | X |
| Reliability | X | X | X | X | X | X |
| Resource utilization | | | | | X | X |
| Reusability | X | | | X | | X |
| Stability | | | | | X | X |
| Suitability | | | | | X | X |
| Supportability | | | X | | X | X |
| Testability | X | X | | | X | X |
| Transferability | | | | | | X |
| Understandability | | X | | | X | X |
| Usability | X | | X | X | X | X |

Comparison among tailored oriented models is more difficult because they use the model in a particular context. The models can be either product oriented (GECUAMO), or for particular domains (Bertoa) or adapted from the point of view of a user (Rawashdeh). Table 3 has been made with almost the same features as the basic models. However it must be noted that the absence of a feature does not invalidate any model.

Table 3 Comparison of Tailored Quality Models

| Characteristic | Bertoa | Gecuamo | Alvaro | Rawashdeh |
|---|---|---|---|---|
| Accuracy | X | | X | X |
| Adaptability | | X | X | |
| Analyzability | | | | |
| Attractiveness | | | | |
| Changeability | X | | X | X |
| Compliance | X | X | X | X |
| Configurability | | | X | |
| Compatibility | | | | X |
| Correctness | | X | | |
| Efficiency | | | X | X |
| Fault Tolerance | | | X | |
| Flexibility | | | | |
| Functionality | X | X | X | X |
| Human Engineering | | | | |
| Installability | | | | |
| Integrity | | | | |
| Interoperability | X | | X | X |
| Learnability | X | X | X | X |
| Maintainability | X | | X | X |
| Manageability | | | | X |
| Maturity | X | X | X | X |
| Modifiability | | | | |
| Operability | X | | | X |
| Performance | | | | |
| Portability | X | | X | |
| Recoverability | X | | | X |
| Reliability | X | | X | X |
| Replaceability | X | | X | |
| Resource utilization | X | X | X | X |
| Reusability | X | | X | |
| Scalability | | | X | |
| Stability | | | X | |
| Security | X | | X | X |
| Self Contained | | | X | |
| Suitability | X | | X | X |
| Supportability | | | | |
| Testability | X | X | X | X |
| Time Behavior | X | | X | X |
| Understandability | X | X | X | X |
| Usability | X | X | X | X |

## 7. Conclusions

The overall conclusion is that there are very general models for assessing software quality and hence they are difficult to apply to specific cases. Also there exist tailored quality models whose range is in small domain, using as starting model the ISO 9126. Models for Free/Open source emphasize the participation of community members.

Tailored Quality Models originated from the Basic Models basic consider a specific domain and selects the features and sub features to consider. The model created in this way is for a specific, particular product or from the point of view of a user domain. Therefore have limitations.

The ISO 9126 model was updated in 2007 by the ISO 25010 that redefines the fundamental characteristics increasing them from six to eight. In the future the developing of models will have to consider these characteristics. Future works will have as main reference this model. In the case of Free Software the aspects of user communities should be considered as a feature of high level because the level of influence in both the construction and the product acceptance.

In all the models studied none has incorporated the aspect of communication as one of the quality factors. At the present time, there is a need for quality components for communications at all levels and especially in complex systems, where it becomes a critical factor because of the Internet.

Finally, we note that in most of the studied models the factors and criteria have the same value which is relative because it depends of the application domain. For example aspects of transferability can be crucial in software that is installed on different machines.

## References

[1] Côte M & Suryn W & Georgiadou E. (2007). ) "In search for a widely applicable and accepted software quality model for software quality engineering paper," *Software Quality Journal*, 15, 401–416

[2] IEEE. (1990). IEEE Std 610.12-1990 (1990)- "IEEE Standard Glossary of Software Engineering Terminology,"
http://web.ecs.baylor.edu/faculty/grabow/Fall2013/csi3374/secure/Standards/IEEE610.12.pdf

[3] IEEE. (1998). Standard for Software Maintenance, Software Engineering Standards *Subcommittee of the IEEE Computer Society*.

[4] Xu Lai & Sjaak Brinkkemper. (2007). "Concepts of Product Software: Paving the Road for Urgently Needed Research," *Technical report, Institute of Information and Computing Sciences, Utrecht University*, The Netherlands. European Journal of Information Systems 16, 531–541.

[5] ISO/IEC IS 9126-1. (2001). Software Engineering - Product Quality – Part 1: Quality Model. *International Organization for Standarization*, Geneva, Switzerland.

[6] Thapar SS & Singh P & Rani S. (2012). "Challenges to the Development of Standard Software Quality Model," *International Journal of Computer Applications* (0975 – 8887) Volume 49–No.10, pp 1-7.

[7] Mc Call, J. A. &, Richards, P. K. & Walters, G. F. (1977). *Factors in Software Quality*, Volumes I, II, and III. US Rome Air Development Center Reports, US Department of Commerce, USA.

[8] Boehm, B. W., Brown, H., Lipow,M. (1978) "Quantitative Evaluation of Software Quality," *TRW Systems and Energy Group*, 1978

[9] Grady, R. B. (1992). *Practical Software Metrics for Project Management and Process Improvement*. Prentice Hall, Englewood Cliffs, NJ, USA

[10] Dromey, R. G. (1995). "A model for software product quality," *IEEE Transactions on Software Engineering*, 21:146-162

[11] ISO/IEC TR 9126-2. (2003). Software Engineering - Product Quality - Part 2: External Metrics. *International Organization for Standardization*, Geneva, Switzerland.

[12] ISO/IEC TR 9126-3. (2003): Software Engineering - Product Quality - Part 3: Internal Metrics, *International Organization for Standardization*, Geneva, Switzerland.

[13] ISO/IEC TR 9126-4. (2004): Software Engineering - Product Quality - Part 4: Quality in Use Metrics. *International Organization for Standardization*, Geneva, Switzerland.

[14] ISO/ IEC CD 25010. (2008). Software Engineering: Software Product Quality Requirements and Evaluation (SQuaRE) Quality Model and guide. *International Organization for Standardization*, Geneva, Switzerland.

[15] Bertoa, M & Vallecillo A. (2002). "Quality Attributes for COTS Components," *I+D Computación*, Vol 1, Nro 2, 128-144.

[16]     Georgiadoui, Elli "GEQUAMO-A Generic, Multilayered, Customizable Software Quality model," *Software Quality Journal*, 11, 4, 313-323. DOI=10.1023/A:1025817312035

[17]     Alvaro A & Almeida E.S. & Meira S.R.L. (2005). "Towards a Software Component Quality Model," *Submitted to the 5th International Conference on Quality Software* (QSIC).

[18]     Rawashdeh A, & Matalkah Bassem. (2006). "A New Software Quality Model for Evaluating COTS Components," *Journal of Computer Science* 2 (4): 373-381, 2006

[19]     Klas Michael & Constanza Lampasona & Jurgen Munch. (2011). "Adapting Software Quality Models: Practical Challenges, Approach, and First Empirical Results," *37th EUROMICRO Conference on Software Engineering and Advanced Applications*, 978-0-7695-4488-5/11, IEEE pp. 341-348

[20]     Ayala Claudia & Hauge, Øyvind & Conradi Reidar & Franch Xavier & Li Jingyue. (2010). "Selection of third party software in Off-The-Shelf-based software development—An interview study with industrial practitioners," *The Journal of Systems and Software*, pp 24-36

[21]     Samarthyam G & Suryanarayana G & Sharma T, Gupta S. (2013). "MIDAS: A Design Quality Assessment Method for Industrial Software," *Software Engineering in Practice*, ICSE 2013, San Francisco, CA, USA, pp 911-920

[22]     Pensionwar Rutuja K & Mishra Anilkumar & Singh Latika. (2013). "A Systematic Study Of Software Quality – The Objective Of Many Organizations, " *International Journal of Engineering Research & Technology (IJERT)*, Vol. 2 Issue 5.

[23]     Trendowicz, A & Punter, T. (2003) "Quality modeling for software product lines," *Proceedings of the 7th ECOOP Workshop on Quantitative Approaches in Object-Oriented Software Engineering, QAOOSE*, Darmstadt, Germany

[24]     Al-Badareen Anas Bassam. (2011). "Software Quality Evaluation: User's View," *International Journal of Applied Mathematics and Informatics*, Issue 3, Volume 5, pp 200-207.

[25]     Dubey, S.K & Soumi Ghosh & Ajay Rana. (2012). "Comparison of Software Quality Models: An Analytical Approach," *International Journal of Emerging Technology and Advanced Engineering*, Volume 2, Issue 2, pp 111-119

[26]     Al-Qutaish, Rafa E. (2010). "Quality Models in Software Engineering Literature: An Analytical and Comparative Study," *Journal of American Science*, Vol. 6(3), 166- 175.

[27]     Ghayathri J & Priya E. M. (2013) "Software Quality Models: A Comparative Study," *International Journal of Advanced Research in Computer Science and Electronics Engineering* (IJARCSEE) ,Volume 2, Issue 1, pp 42-51.

[28]     Samadhiya Durgesh & Wang Su-Hua & Chen Dengjie.(2010), "Quality Models: Role and Value in Software Engineering," *2nd International Conference on Software Technology and Engineering(ICSTE)*. Pp 320-324.

[29]     ASQ (2007). *American Society for Quality. Glossary*. http://www.asq.org/glossary/q.html , jan 2007.

[30]     Alvaro A. &. Almeida E.S & Meira. S.R.L (2010). "A Software Component Quality Framework," *ACM SIGSOFT SEN 35, 1* (Mar. 2010), 1-4.

[31]     Glott R. & Arne-Kristian Groven & Kirsten Haaland & Anna Tannenberg. (2010). "Quality models for Free/Libre Open Source Software– towards the "Silver Bullet"?," *EUROMICRO Conference on Software Engineering and Advanced Applications IEEE Computer Society*, 439-446.

[32]     Adewumi Adewole, Sanjay Misra and Nicholas Omoregbe. (2013). "A Review of Models for Evaluating Quality in Open Source Software," *2013 International Conference on Electronic Engineering and Computer Science, IERI Procedia* 4, 88 – 92.

[33]     Haaland K & Groven AK & Regnesentral N & Glott R & Tannenberg A. (2010). "Free/Libre Open Source Quality Models-a comparisonbetween two approaches," *4th FLOS International Workshop on Free/Libre/Open Source Software*, pp. 1-17.

[34]     Duijnhouwer FW & Widdows. (2003). "C. Open Source Maturity Model, ". *Capgemini Expert Letter.*

[35]     Wasserman AI & Pal M & Chan C. (2006). "Business Readiness Rating for Open Source," *Proceedings of the EFOSS Workshop*, Como, Italy.

[36]     Samoladas I & Gousios G & Spinellis D & Stamelos I. (2008). "The SQO-OSS quality model: measurement based open source software evaluation," *Open source development, communities and quality*. 237-248.

[37]     AL-Badareen Anas Bassam & Mohd Hasan Selamat & Marzanah A. Jabar & Jamilah Din & Sherzod Turaev. (2011). "Software Quality Models: A Comparative Study", *J.M. Zain et al. (Eds.): ICSECS 2011, Part I, CCIS* 179, pp. 46–55. © Springer-Verlag Berlin Heidelberg.